\documentclass[prd,aps,showpacs,nofootinbib,preprint]{revtex4-1}
\usepackage{graphicx,color,amsmath,amsxtra}
\usepackage{epsf}
\usepackage{amssymb}
\usepackage{enumerate}
\usepackage{hhline}
\usepackage{array}
\usepackage{tabularx}
\usepackage{hangcaption}

\newcommand{\newc}{\newcommand}
\newc{\be}{\begin{equation}}
\newc{\ee}{\end{equation}}
\newc{\beq}{\begin{eqnarray}}
\newc{\eeq}{\end{eqnarray}}

\begin{document}
\rightline{\tt{MIT-CTP-4209}}

\title{Localizing fields on brane in magnetized backgound
}
\author{
Shih-Hao Ho\footnote{E-mail address: shho@mit.edu}
}
\affiliation{Center for Theoretical Physics, Massachusetts Institute of Technology, Cambridge, Massachusetts 02139, USA}
\author{
C. Q. Geng\footnote{E-mail address: geng@phys.nthu.edu.tw}
}
\affiliation{Department of Physics, National Tsing Hua University, Hsinchu, Taiwan 300\\
National Center for Theoretical Sciences, Hsinchu, Taiwan 300
}


\begin{abstract}
To localize the scalar, fermion, and abelian gauge fields 
on our 3-brane, 
a simple mechanism 
with a hypothetical ``magnetic field'' in the bulk is proposed. 
This mechanism is to treat all fields in the equal footing without ad hoc consideration.
In addition,  the machanism can be easily realized in a flat dimension six Minkowski  space
and it works even in the weak coupling limit.

\end{abstract}


\maketitle

It is known that 
the  brane-world scenario allows us to use infinite extra dimensions without compactification. Since the low energy effective theory should be four-dimensional, the mechanism to localize matter fields naturally on a brane  still is a main topic
 in the brane-world scenario~\cite{Shifman:2009df}.
In Ref.~\cite{Rubakov:1983bb}, Rubakov and Shaposhinikov proposed to localize fields on our 3-brane by topological defects\footnote{In Ref. ~\cite{Palma:2005xv}, the author attempted to reproduce SM phenomenology exploiting the type of fermion local- 
ization by the mechanism introduced in Ref. ~\cite{Rubakov:1983bb}
}. 
The mechanism 
to localize scalar and fermion 
is essentially 
given in Refs.~\cite{Dashen:1974cj,Jackiw:1975fn}. 
The feature of the mechanism is the existence of zero modes,  identified to be our standard model fields. 
In the literature, there have been many proposals  to trap fields in our four-dimesional world \cite{Shifman:2009df}. 
In this  paper, we try to illustrate a simple and old way to localize particles on a plane by applying a strong magnetic field. 


Let us consider  the Lagrangian in a theory with a massive complex scalar field $\phi$,
and a (classical) background gauge field $a_M$ associated with a $U(1)_X$ in a flat dimension six spacetime,\footnote{Our convention is  $\eta^{MN}=diag(+1,-1,-1,-1,-1,-1)$, while Greek and Latin indices  represent the (3+1) spacetime and   all the dimensions indices, respectively.}
\beq \label{scalar lagrangian}
\mathcal{L}_S=(\partial_M+i g a_M)\phi^* (\partial^M-ig a^M)\phi - m_s^2\phi^*\phi\,.
\eeq
%
The equation of motion for these fields can be easily derived to be
\beq  \label{eom s}
&&\left[\partial_M\partial^M- i g (\partial_M a^M)- 2 i g a_M \partial^M - g^2 a_M a^M +m_s^2\right]\phi=0\,.
\eeq

%
For the case in which  ``magnetic field'' is present along the extra dimensions, i.e. $f_{45}=-f_{54}\equiv B$, $f_{\mu\nu}=0$ for $\mu$,$\nu = 0,1,2$ and 3
in the gauge $a_M=(0,0,0,0,-B x_5/2,B x_4/2)$ with $f_{MN}\equiv \partial_M a_N - \partial_N a_M$,\footnote{The wavefunctions $depend$ on the gauge choice and they differ only by a phase factor as expected \cite{Haugset}.}
 Eq.(\ref{eom s}) becomes \footnote{Here we note the location of this 3-brane along the extra dimensions is $arbitrary$ and it also depends on the gauge choice of classical field $a_M$. For instance, the choice of $a_M(x)= (0,0,0,0,-B(x_5+a)/2, B(x_4+b)/2)$ with some constants $a$, $b$ will give us the same background ``magnetic field" but it will also change the location of the brane from $x_4=x_5=0$ to $x_4=-b$, $x_5=-a$. But the important property is that the wavefunctions and the propagators fall off exponentially outside the brane.}
\beq \label{eom s1}
\left[\eta^{\mu\nu}\partial_{\mu}\partial_{\nu}+(i\partial_4-\frac{gB}{2}x_5)^2+(i\partial_5+\frac{gB}{2}x_4)^2+m_s^2\right]\phi=0\,.
\eeq
Clearly, it reduces to a non-relativistic quantum mechanics problem. Using the raising and lowering operator method, we  obtain the energy spectrum and ground state wave function:
%
%
\beq
&& p_0^2=p_1^2+p_2^2+p_3^2+(2n+1)|gB|+m_s^2 \ , n=0,1,2,\cdots\\
&& \label{eom s01}
\phi_{n=0}=
N_s e^{ip_{\mu}x^{\mu}} r^l e^{i l\theta} e^{-|gB|r^2/4} \ , l=0,1,2,\cdots\,,
\eeq 
%
representing the Landau levels for spin-zero particles with
 the lowest Landau level (LLL) for n=0 
of 
the ground state energy  $E_0^2=p_1^2+p_2^2+p_3^2+|gB|+m_s^2$,
where $N_s$ is the normalization constant, $r^2=x_4^2+x_5^2$ and $tan \theta =x_5/x_4$. This shows that the ground state with respect to the bulk dynamics is localized near $r^2 \equiv r_0^2 = 2 l/|gB|$ as long as the bulk magnetic field is large enough. However, we observe
that, in the presence of the term $|gB|$, these ``zero modes'' \footnote{
The ``zero mode" referred by us means ``zero mode of Landau level", i.e. the Lowest Landau Level (LLL).} carry information from extra dimensions even if the bulk mass is absent.

We emphasis that the "center" of wavefunction in Eq.(\ref{eom s01}) is $r=r_0$, which $depends$ on the angular quantum number $l$. Hence the field is concentrated at $r=r_0(l)$ with characteristic length $1/|gB|$ for a fixed $l$\footnote{This can be seen more clearly in Landau gauge.}, we cannot argue which value of $l$ is a preferred one since all values of $l$ correspond to the LLL. The absolute position in extra dimensions, however, is irrelevant in our consideration, the relevant quantity is always the relative positions of different fields along the extra dimensions. We believe this will become much clearer when the interactions between fields are taken into account. For present purpose, we emphasis the $dynamics$ of fields is effectively confined on 3-brane even if we consider the free fields in an magnetized background. This can be seen as we consider the Green's function in the second part of this paper.



We now turn to the fermion case, the goal is still to find a zero mode localized on our brane~\cite{Jackiw:1984ji}. The Lagrangian for a massive Dirac fermion $\Psi$ on a background gauge field $a_M$ in a flat dimension six spacetime is
\beq \label{dirac lagrangian}
\mathcal{L}_D=\bar{\Psi}(i D_{M}\Gamma^{M}-m_f)\Psi\,,
\eeq
where $D_M\equiv \partial_M-ig a_M$ with the coupling constant g and 
the representations of 8 by 8 Dirac matrices are chosen to be:
\beq \label{gamma}
\Gamma^{\mu}=\gamma^{\mu} \otimes \sigma_1 = \left(\begin{array}{cc}0 & \gamma^{\mu} \\\gamma^{\mu} & 0\end{array}\right), &&
~~\Gamma^4=i\gamma^5 \otimes\sigma_1=\left(\begin{array}{cc}0 & i \gamma^{5} \\\ i \gamma^{5} & 0\end{array}\right), \nonumber \\
 \Gamma^5=\textsc{1}\otimes i \sigma_2=\left(\begin{array}{cc}0 & 1 \\\ -1 & 0\end{array}\right), && ~~ \Gamma^7=\Gamma^0\Gamma^1\Gamma^2\Gamma^3\Gamma^4\Gamma^5= \left(\begin{array}{cc}1 &0\\ 0& -1\end{array}\right), 
\eeq
with the four dimensional gamma matrices  in the chiral representation. The gamma matrices in Eq.~(\ref{gamma})
satisfy the Clifford algebra 
\beq 
\{\Gamma^M ,\Gamma^N\} &= & 2 \eta^{MN}, \ \eta^{MN}=diag(+1,-1,-1,-1,-1,-1), \\
\{\Gamma^{M}, \Gamma^7 \} &=&0, \ (\Gamma^7)^2=1,
\eeq with M,N =$0,1,2,3,4,5$.
%
By defining the 6D chiral states for the eight-componet spinor $\Psi$:
\beq
\Psi = \left(\begin{array}{c}\Psi_+ \\ \Psi_-\end{array}\right), with  \ \Gamma^7\Psi_{\pm}=\pm \Psi_{\pm}\,,
\eeq
we can decompose equations of motion for $\Psi_{\pm}$ into
\beq
&& (i\gamma^{\mu}D_{\mu}-\gamma^5 D_4+i \textbf{1} D_5)\Psi_-=m_f \Psi_+ \,, \\
&& (i\gamma^{\mu}D_{\mu}-\gamma^5 D_4-i \textbf{1} D_5)\Psi_+=m_f \Psi_- \,.
\eeq
%
Consequently, 
these equations can be rewritten as
\beq
&& (i\gamma^{\mu}D_{\mu}-\gamma^5 D_4-i \textbf{1} D_5)(i\gamma^{\mu}D_{\mu}-\gamma^5 D_4+i \textbf{1} D_5)\Psi_-=m_f^2 \Psi_-\,, \\
&& (i\gamma^{\mu}D_{\mu}-\gamma^5 D_4+i \textbf{1} D_5)(i\gamma^{\mu}D_{\mu}-\gamma^5 D_4-i \textbf{1} D_5)\Psi_+=m_f^2 \Psi_+
\,,
\eeq 
which lead to
%
\beq
&&\left[\eta^{\mu\nu}\partial_{\mu}\partial_{\nu}+(i\partial_4-\frac{gB}{2}x_5)^2+(i\partial_5+\frac{gB}{2}x_4)^2+gB \gamma^5+m_f^2\right]\Psi_-=0 \label{eom f1}\,, \\
&&\left[\eta^{\mu\nu}\partial_{\mu}\partial_{\nu}+(i\partial_4-\frac{gB}{2}x_5)^2+(i\partial_5+\frac{gB}{2}x_4)^2-gB \gamma^5+m_f^2\right]\Psi_+=0 \,.\label{eom f2}
\eeq 
Since  $\gamma^5$ commutes with the whole operator in the brackets in  Eqs.~(\ref{eom f1}) and (\ref{eom f2}), 
it
can be replaced by the eigenvalue $\gamma$ which is the chirality of a 4D fermion: 
\beq
\Psi_{\pm}=\left(\begin{array}{c}\Psi_{\pm_L} \\\Psi_{\pm_R} \end{array}\right), \ \gamma^5 \Psi_{\pm_{L,R}}=\gamma \Psi_{\pm_{L,R}}, 
\eeq with $\gamma=-1 (+1)$ for L(R).
We  see that both Eqs.~(\ref{eom f1}) and  (\ref{eom f2}) take the same form as Eq. (\ref{eom s1}).
Therefore, we can write down the energy spectrum and ground state wavefunction immediately:
\beq
&& p_0^2=p_1^2+p_2^2+p_3^2+(2n+1+sgn(gB)\gamma)|gB|+m_f^2 \ , n=0,1,2,\cdots\ \  for\  \Psi_- \label{eom f3}\\
&& p_0^2=p_1^2+p_2^2+p_3^2+(2n+1-sgn(gB)\gamma)|gB|+m_f^2  \ , n=0,1,2,\cdots\ \   for \  \Psi_+ \label{eom f4}\\
&&\Psi_{\pm_{n=0}}=
N_f e^{ip_{\mu}x^{\mu}} r^l e^{i l\theta} e^{-|gB|r^2/4} \ \omega_{\gamma} \ , l=0,1,2,3.... 
\eeq 
where $N_f$ is the normalization constant, $\omega_{\gamma}$ denotes the eigenvector of $\gamma^5$, $r^2=x_4^2+x_5^2$ and $tan \theta =x_5/x_4$ as mentioned early. Now, we observe that for the lowest energy state with $n=0$ and $\gamma=\pm sgn(gB)$
 for $\Psi_\pm$, 
  the energy is $E_0^2=p_1^2+p_2^2+p_3^2$ if the bulk mass is absent. 
Clearly,  there exist two zero modes with respect to the bulk dynamics, corresponding to $\Psi_{-L}$ and $\Psi_{+R}$ if 
$sgn(gB)>0$.

Next, we consider the simplest case of complex massive vector gauge fields $A_M$ associated with  $U(1)_A$,
 which also carry a charge of $U(1)_X$, i.e. $A_M \rightarrow A_M+\partial_M \Lambda(x)$ and $A_M \rightarrow e^{i \alpha(x)} A_M$ under $U(1)_A$ and $U(1)_X$ gauge transformations, respectively. Based on Ref.~\cite{Jackiw:1997vw}, we can construct a gauge invariant Lagrangian under these two $U(1)$ groups with the help of a scalar Stueckelberg field $\rho$ whose transformation laws are $\rho \rightarrow \rho+ m_v \Lambda^{\prime}(x)$ and $\rho \rightarrow e^{i \alpha(x)} \rho$ under $U(1)_A$ and $U(1)_X$, respectively \cite{Ruegg:2003ps}. Then, the Langrangian reads:
\beq \label{v}
\mathcal{L}_V=\frac{-1}{2}\mathbb{F}^{*}_{MN}\mathbb{F}^{MN}-ig f^{NM}V^*_N V_M+m_v^2 V^*_M V^M-\frac{1}{\xi}|D_M A^M+ \xi m_v\rho|^2\,,
\eeq
where $V_M \equiv A_M-\frac{1}{m}D_M \rho$, $\mathbb{F}_{MN} \equiv D_M A_N-D_N A_M$ and $\xi$ is the gauge parameter. The gauge invariance of $U(1)_X$ is obvious and the invariance under $U(1)_A$ is guaranteed by the nontrivial condition of $D_M \Lambda^{\prime} =\partial_M \Lambda$.\footnote{Note that 
$(D_M D^M+\xi m^2)\Lambda^{\prime}=0$.} 
We can derive equations of motion by varying $A_M$ and $\rho$, given by
\beq
(D^2 \eta^{MN} && -D^M D^N)A_M +D_M(\frac{i g}{m_v} f^{MN}\rho)- i  f^{NM}A_M+\frac{i g}{m_v}f^{NM}D_M \rho \nonumber \\
&&+m_v^2(A^N-\frac{1}{m_v}D^N \rho) +\frac{1}{\xi}(D^N D^M A_M+m_v D^N \rho)=0 \label{v1} 
\eeq and 
\beq 
&&D_N \left( \frac{1}{m_v}D_M \mathbb{F}^{MN}-\frac{i g}{m_v} f^{NM}(A_M -\frac{1}{m_v}D_M \rho)+m_v(A^N-\frac{1}{m_v}D^N \rho)\right)-m_v^2\rho =0 \label{v2} \,.
\eeq
We will concentrate on the case with
$f^{MN}=$constant and  choose the Feynman gauge $\xi=1$. 
Consequently,  Eq.(\ref{v1}) is reduced to
\footnote{  In the original Stuckelberg formalism, the vector field $A_M$ and scalar field $\rho$ satisfy the same field equation \cite{Ruegg:2003ps}:
  \beq 
  &&(\partial_M \partial^M+m^2)A_M =0 \nonumber \\
  &&(\partial_M \partial^M+m^2) \rho=0 \nonumber 
  \eeq
  In our consideration, the field equations are not the same owing to an external background field. We can derive the equation of motion for 
  scalar field $\rho$ from Eq.~(\ref{v2}) by using the relation $[D_M , D_N]= - i g f_{MN}$:
  \beq
  (D_{M}D^{M}+m^2)\rho=0.
  \eeq
}
\beq
\left[(D^2 +m_v^2)\eta^{MN}-2ig f^{MN}\right] A(x) \epsilon_{M}=0\label{v3}\,.
\eeq
Here, we still have degree of freedoms to set $A_0=0$ and write $A_M \equiv A(x) \epsilon_M$
with the polarization vector $\epsilon_M$ 
%
\beq
\epsilon_{\mu}=\left(\begin{array}{c}\delta_{\mu 1} \\\delta_{\mu 2} \\\delta_{\mu 3} \\0 \\0\end{array}\right), \  \mu=1,2,3;  \  \epsilon_{\pm}= \frac{1}{\sqrt{2}}\left(\begin{array}{c}0 \\0 \\0 \\1 \\\pm i\end{array}\right)\,.
\eeq
In this basis, Eq. (\ref{v3}) becomes diagonal, given by
%
\beq
\left[(D^2 +m_v^2)\eta^{MN}+2 g B \chi g^{MN}\right] A(x) \epsilon_{M}=0 \label{v4}\,,
\eeq 
where $\chi=0$ for $M=1,2,3$ and $\chi=\mp 1$ for $M=\pm$. 
Hence, the energy spectrum and coordinate space component of $A_M$ is the solution of Eq.(\ref{eom s1}) with the replacement $m^2 \rightarrow m^2+2gB\chi$, found to be
\beq
&& p_0^2=p_1^2+p_2^2+p_3^2+(2n+1+2 sgn(gB)\chi)|gB|+m_v^2 \ , n=0,1,2,\cdots\\
&& A(x)_{n=0}=
N_v e^{ip_{\mu}x^{\mu}} r^l e^{i l\theta} e^{-|gB|r^2/4} \ , l=0,1,2,\cdots\,,
\eeq where $N_v$ is the normalization constant, $r^2=x_4^2+x_5^2$ and $tan \theta =x_5/x_4$.
The ground state energy is $E_0^2=p_1^2+p_2^2+p_3^2+(1+2 sgn(gB)\chi)|gB|+m_v^2$.  Therefore, in the viewpoint of 4D, we have a localized massive vector field $A_{\mu}$ with an effective mass $\sqrt{|gB|}$ in 4D and two localized scalar field $A_{\pm}$ which carry an extra degree of freedom with respect to the extra dimensional ``handness'' with effective squared masses of
 $m_v^2+(1+2sgn(gB))|gB|$ and $m_v^2+(1-2sgn(gB))|gB|$ for ``right-handed'' and ``left-handed'' states, respectively.

There are three possible cases of $A_{\pm}$, which are listed  in Table~I.
\begin{table}[htdp] \label{t}
\begin{center}
\caption{Types of zero modes for $A_{\pm}$}
\begin{tabular}{|c|c|c|} 
\hline
Mass parameter  & Effective masses of zero modes & Type of modes \\
\hline
$m_v^2>|gB|$ & $m_{eff}^2|_{\chi=\pm 1}=m_v^2+(1\mp 2sgn(gB))|gB|>0$ & $A_{\pm}$ are both massive \\
\hline
$m_v^2=|gB|$, $sgn(gB)>0$ & $m_{eff}^2|_{\chi=+1}=4|gB|$, $m_{eff}^2|_{\chi=-1}=0$ & $A_+$ massless, $A_-$ massive \\
$m_v^2=|gB|$, $sgn(gB)<0$ & $m_{eff}^2|_{\chi=-1}=4|gB|$, $m_{eff}^2|_{\chi=+1}=0$ & $A_-$ massless, $A_+$ massive  \\
\hline
$m_v^2<|gB|$, $sgn(gB)>0$ & $m_{eff}^2|_{\chi=+1}>0$, $m_{eff}^2|_{\chi=-1}<0$ & $A_+$ tachyonic mode, $A_-$ massive \\
$m_v^2<|gB|$, $sgn(gB)<0$ & $m_{eff}^2|_{\chi=-1}>0$, $m_{eff}^2|_{\chi=+1}<0$ &  $A_-$ tachyonic mode, $A_+$ massive \\
\hline
\end{tabular}
\end{center}
\label{T1}
\end{table}%
 We note that in the third case, there is a heavy scalar for $\chi=-sgn(gB)$ and a unstable mode for $\chi=sgn(gB)$. This is the so-called zero mode problem and a general feature when we consider the spectrum  of a vector field in the presence of magnetic field~\cite{Nielsen:1978rm, Hashimoto:1997gm}.\footnote{In Ref.~\cite{Hashimoto:1997gm},  a  gauge theory with U(2) on a $T^2$ has been considered. There exists an instability in this guage theory, which corresponds to the instability in a brane-anti-brane configuration.} We also note that as a special case, to have a massless vector field $A_{\mu}$ in 4D, we may 
 re-consider  the Lagrangian in Eq.~(\ref{v}) with
 an ``imaginary'' mass term as 
\beq
\mathcal{L}_V=\frac{-1}{2}\mathbb{F}^{*}_{MN}\mathbb{F}^{MN}-ig f^{NM}V^*_N V_M-\frac{1}{\xi}|D_M A^M+m_v\rho|^2+(i m_v^{\prime})^2 V^*_M V^M\,,
\nonumber
\eeq 
where  $(m^{\prime}_v)^2=|gB|$.


Despite of the localized wavefunctions, we 
emphasize that 
the effective dynamics is confined on the 3-brane when the strong background field (and massless limit) is applied. This effective dimensional reduction can be seen in either the coordinate/momentum space reduction~\cite{Gusynin:1999pq} or the phase space reduction~\cite{Dunne:1989hv}. The phase space reduction is examined in detail in Ref. \cite{Dunne:1989hv}. Here, we would like to see the propagators for scalar and fermion fields and show that their propagations are indeed localized on the brane with the characteristic length $l_B^{-2} \equiv |gB|$.


Using Schwinger's proper time method \cite{Schwinger:1951nm}, we can calculate the propagators of scalar and fermion fields in the presence of a constant background. Since the presence of the constant background field breaks the translational invariance, we recover the translation invariance by factoring out the Schwinger phase:
\beq
&&G_s(x,y)=e^{i g \int^x_y a_M(\xi_M) d\xi^M } \tilde{G}_s(x-y) \,,\\
&&G_f(x,y)=e^{i g \int^x_y a_M(\xi_M) d\xi^M } \tilde{G}_f(x-y)\,,
\eeq 
where $G_s$ and $G_f$ stand for the propagators for scalars and fermions, respectively, and the path of the  integral is a straight line.

The translational parts of the propagators $\tilde{G}_s$ and $\tilde{G}_f$ in the momentum space are given by
\beq \label{sp}
\tilde{G_s}(p)&=&\int d^6x \tilde{G_s}(x) e^{ipx}
=\int^{\infty}_0 ds e^{-is(m^2+p_{\mu}^2+\frac{tan(|gB|s)}{|gB|s}(p_4^2+p_5^2))} \frac{1}{cos(|gB|s)}\,,\\
%
\label{fp}
\tilde{G_f}(p)&=&\int d^6x \tilde{G_f}(x) d^{ipx} 
=\int^{\infty}_0 ds e^{-is(m^2+p_{\mu}^2+\frac{tan(|gB|s)}{|gB|s}(p_4^2+p_5^2))}\times
\nonumber\\
&& 
[(p_{\mu} \Gamma^{\mu}+m)(1+sgn(gB)\Gamma^4\Gamma^5 tan(|gB|s))
-(p_4\Gamma^4+p_5\Gamma^5)(1+tan^2(|gB|s)],\
\eeq
%
which can be decomposed  to be
 \cite{Gusynin:1999pq,Chodos:1990vv}:
\beq
&&\tilde{G_s}(p)=i e^{-\frac{p_{\bot}^2}{|gB|}} \sum_{n=0}^{\infty}(-1)^n \frac{L_n(x)}{p_{\mu}^2-m^2-(2n+1)|gB|}\,,   \label{sp1}\\
&&\tilde{G_f}(p)=i e^{-\frac{p_{\bot}^2}{|gB|}} \sum_{n=0}^{\infty}\frac{(-1)^n D_n(gB,p)}{p_{\mu}^2-2|gB|n-m^2}\,,    \label{fp1}
\eeq
respectively, with
\beq
D_n(gB,p)=&&(p_{\mu}\Gamma^{\mu}+m) \left[(1+i \Gamma^4 \Gamma^5)L^0_n(x)-(1+i\Gamma^4\Gamma^5)L^0_{n-1}(x)\right] \nonumber \\
&&+4(p_4\Gamma^4+p_5\Gamma^5) L^1_{n+1}(x)\,,
\eeq
where $p_{\bot}^2\equiv p_4^2+p_5^2$,  $x \equiv \frac{2(p_4^2+p_5^2)}{|gB|}$ and  $L_n^{(\alpha)}(x)$ are the (associated) Laguerre polynomials.
In the LLL for $n=0$, the propagators take the forms:
\beq
&&\tilde{G_s}(p)=i e^{-\frac{p_{\bot}^2}{|gB|}} \frac{1}{p_{\mu}^2-m^2-|gB|} \label{sp2}\,,\\
&&\tilde{G_f}(p)= i e^{-\frac{p_{\bot}^2}{|gB|} } \frac{p_{\mu}\Gamma^{\mu}+m}{p_{\mu}^2-m^2}(1-i\Gamma_4\Gamma_5) \label{fp2}\,,
\eeq
%
%
which  are the exact free propagators in (3+1) dimension in the region of $p_{\bot}^2<<|gB|$. 
We can also transform them back to the coordinate space:
\beq
&&\tilde{G}_s(x-y)=\frac{1}{4\pi l_B^2}e^{-\frac{(x_4-y_4)^2+(x_5-y_5)^2}{4 l_B^2}} g_s(x_{\mu}-y_{\mu})\,, \\
&&\tilde{G}_f(x-y)=\frac{1}{4\pi l_B^2}e^{-\frac{(x_4-y_4)^2+(x_5-y_5)^2}{4 l_B^2}} g_f(x_{\mu}-y_{\mu})(1-i\Gamma_4\Gamma_5)\,,
\eeq
where  $g_s(x_{\mu})$ and $g_f(x_{\mu})$ are the ordinary scalar and fermion propagators in (3+1) spacetime, given by
\beq
&& g_s(x_{\mu}-y_{\mu})=\int \frac{d p^4}{(2\pi)^4} e^{-i p_{\mu}(x-y)^{\mu}}\frac{1}{p_{\mu}^2-m^2-|gB|}\,,  \\
&& g_f(x_{\mu}-y_{\mu})=\int \frac{d p^4}{(2\pi)^4} e^{-i p_{\mu}(x-y)^{\mu}}\frac{p_{\mu}\Gamma^{\mu}+m}{p_{\mu}^2-m^2}\,.
\eeq
%
It is clear that the propagations along the extra dimension are indeed suppressed by the Gaussian function with the characteristic length $l_B$. 

To consider in Feynman gauge $\xi=1$,  we start from the differential equation 
\beq
\left[(D^2 +m_v^2)\eta_{MN}-2ig f_{MN}\right] G^{NP}_v(x,y)=i \delta^P_M \delta^{(6)}(x-y)   \label{v5}\,.
\eeq
By applying a rotation matrix $\mathbf{P}$ with
\beq
(\mathbf{P})_M^N=\left(\begin{array}{cccccc}1 & 0 & 0 & 0 & 0 & 0 \\0 & 1 & 0 & 0 & 0 & 0 \\0 & 0 & 1 & 0 & 0 & 0 \\0 & 0 & 0 & 1 & 0 & 0 \\0 & 0 & 0 & 0 & \frac{1}{\sqrt{2}} & \frac{1}{\sqrt{2}} \\0 & 0 & 0 & 0 & \frac{i}{\sqrt{2}} & \frac{-i}{\sqrt{2}}\end{array}\right)\,,
\eeq
Eq.(\ref{v5}) in the matrix form becomes 
\beq
i \mathbf{1} \delta^{(6)} (x-y)&&=\mathbf{P}^{-1} \left[(D^2 +m_v^2)\mathbf{\eta}-2ig \mathbf{f}\right]\mathbf{P} \mathbf{P}^{-1} \mathbf{G}\mathbf{P}  \nonumber \\
&&=\left[D^2+m_v^2+2gB \chi\right]\mathbf{\eta} \cdot \mathbf{\hat{G}}\,,
\eeq  
where $\mathbf{\hat{G}}\equiv \mathbf{P}^{-1} \mathbf{G}\mathbf{P} $ and $\chi=0,-1,1$ for $M=0\sim3, 4,5$, respectively. 
It is clear that we can obtain
the scalar propagator again with the matrix structure $g_{MN}$ and the replacement $m_s^2\rightarrow m_v^2+2gB\chi$.
 The translational invariant part of the propagator for the vector field (up to a Schiwinger phase) is given by:
\beq
\tilde{G}_{MN}(p)=\mathbf{P}_M^P\hat{G}_{PQ}(p)(\mathbf{P^{-1}})^Q_N=\mathbf{P}_M^P \left[\eta_{PQ}\times \tilde{G}_s(p)|_{m_s^2\rightarrow m_v^2+2gB\eta} \right](\mathbf{P^{-1}})^Q_N \,,
\eeq where $G_s(p)$ is given in Eq.~(\ref{sp}).
%

Some remarks are in order. First, we emphasize
 that we can take the limit $g\rightarrow 0$ but keep the combination $|gB|>>p^2_{\bot}$
 to guarantee  
 the use of the LLL. Secondly, as shown explicitly in Eqs.~(\ref{sp2}) and (\ref{fp2}), when we consider a scattering amplitude, the momentum integral is effectively four-dimensional since the transverse momentum integral is factored out and becomes a Gaussian-type integral which is only a finite constant. This is still true even if we go beyond the LLL as we can see from Eqs.~(\ref{sp1}) and (\ref{fp1}). 

In summary, we have considered a simple way to localize scalar, fermion and abelian gauge fields on our 3-brane. The advantage of this consideration is that it is a simple and well-studied setup. We  note that the remnant from the ``magnetic field'' gives  effective mass terms for scalars even if we set the the bulk mass to zero, leading to an effective massive scalar theory in (3+1) spacetime. On the other hand, if we start from a massless bulk fermion theory, the effective theory in (3+1) spacetime will still be massless. 

To localize a massive gauge field, an additional Stuckelberg field is needed to preserve the gauge invariance. The choice of this bulk mass term is crucial for us to avoid the ``zero-mode'' problem. However, if we would like to reduce to an effective massless abelian gauge theory in (3+1) spacetime, an imaginary mass could be considered, leading to
a tachyonic mode in the theory. This unavoidable tachyonic mode shall have interesting phenomenology 
and deserve further studies.

\section*{Acknowledgments}

We are grateful to Professor Roman Jackiw for many useful discussions and reading the manuscript. We would like to thank Yuhsin Tsai for helpful discussions.
The work by S.H.H. is supported by
the National Science Council of R.O.C. under
Grant number: NSC98-2917-I-564-122 and that 
 by C.Q.G. is supported in part by 
the National Science Council of R.O.C. under
Grant number: NSC-98-2112-M-007-008-MY3.


\end{document}